\documentclass[12pt]{article}
\usepackage{graphicx}
\usepackage{cite}
\usepackage{color}
\newcommand{\bm}[1]{\mbox{\boldmath $#1$}}
\newcommand{\fnd}[2]{\frac{\textstyle #1}{\textstyle #2}}
\newcommand{\abs}[1]{\left| #1\right|}
\newcommand{\fndrs}[4]{\fnd{\raisebox{#1}{$#2$}}{\raisebox{#3}{$#4$}}}
\newcommand{\x}[1]{{\textstyle #1}}
\newcommand{\xrm}[1]{{\textstyle \mbox{\rm #1}}}
\newcommand{\dissum}[2]{\displaystyle \sum_{#1}^{#2}}
\begin{document}
\title{Possible $\psi(5S)$, $\psi(4D)$, $\psi(6S)$ and $\psi(5D)$
signals in $\Lambda_{c}\bar{\Lambda}_{c}$}
\author{
Eef~van~Beveren, Xiang~Liu, Rita~Coimbra,\\ [10pt]
{\small\it Centro de F\'{\i}sica Computacional,
Departamento de F\'{\i}sica,}\\
{\small\it Universidade de Coimbra, P-3004-516 Coimbra, Portugal}\\
{\small\it eef,liuxiang,rita@teor.fis.uc.pt}\\ [10pt]
\and
and George~Rupp\\ [10pt]
{\small\it Centro de F\'{\i}sica das Interac\c{c}\~{o}es Fundamentais,}\\
{\small\it Instituto Superior T\'{e}cnico, Universidade T\'{e}cnica de
Lisboa,}\\
{\small\it Edif\'{\i}cio Ci\^{e}ncia, P-1049-001 Lisboa, Portugal}\\
{\small\it george@ist.utl.pt}\\ [.3cm]
{\small PACS number(s): 14.40.Gx, 13.60.Rj, 14.20.Lq, 11.80.Gw}
}


\maketitle

\begin{abstract}
It is shown that the $\Lambda_{c}^{+}\Lambda_{c}^{-}$ signal
recently reported by the Belle Collaboration \cite{PRL101p172001}
contains clear signs of the $\psi(5S)$ and the $\psi(4D)$
$c\bar{c}$ vector states, and also some indication
for the masses and widths of the $\psi(6S)$ and $\psi(5D)$.
Moreover, it is argued that the threshold behaviour
of the $\Lambda_{c}^{+}\Lambda_{c}^{-}$ cross section
suggests the presence of the hitherto undetected $\psi(3D)$ state
not far below the $\Lambda_{c}^{+}\Lambda_{c}^{-}$ threshold.
\end{abstract}

Very recently \cite{PRL101p172001}, the Belle Collaboration announced
the observation of a near-threshold enhancement
by studying the $e^{+}e^{-}\to\Lambda_{c}^{+}\Lambda_{c}^{-}$ cross section.
The experimental analysis resulted in a mass and width
of this enhancement of $M=(4634^{+8}_{-7})$(stat.)$^{+5}_{-8}$(sys.)~MeV
and $\Gamma_{\mathrm{tot}}=92^{+40}_{-24}$(stat.)$^{+10}_{-21}$(sys.)~MeV,
respectively \cite{PRL101p172001}, with a significance of $8.8$ $\sigma$.
A peculiar aspect of this new experimental observation
is that the main signal lies close to
the $\Lambda_{c}^{+}\Lambda_{c}^{-}$ threshold,
making an understanding of this structure a highly topical issue.
One of the aims of the present paper is to demonstrate that it
is not difficult to explain the new enhancement within the framework of the
Resonance-Spectrum-Expansion (RSE) model \cite{IJTPGTNO11p179}, though not as
a new resonance, but rather a peaked structure resulting from the opening of
the $\Lambda_{c}^{+}\Lambda_{c}^{-}$ threshold
and a nearby zero in the amplitude.

Another important point of the Belle observation
is that it amounts to the first measurement of the production
of a pair of charmed baryons directly created in an $e^{+}e^{-}$ collision
experiment. This opens up a new window to further understand
the mass spectrum of highly excited $c\bar c$ states.
In the following, besides exploring the structure
near the $\Lambda_{c}^{+}\Lambda_{c}^{-}$ threshold,
we shall try to extract useful information on $c\bar{c}$ vector states
from the measured cross section of the process
$e^{+}e^{-}\to\Lambda_{c}^{+}\Lambda_{c}^{-}$.

The discovery of charm was announced in 1974,
after the observation of the $c\bar{c}$ vector meson
$J$ at BNL \cite{PRL33p1404} and $\psi$ at SLAC \cite{PRL33p1406},
which was then baptised ``$J/\psi$'' or, alternatively, ``$\psi (1S)$''.
The subsequent discovery of further charmonium vector states, i.e.,
the $\psi (2S)$ \cite{PRL33p1453,PRL35p1124} and
the $\psi (1D)$ \cite{PRL39p526} at SLAC,
and three more peaks observed by the DASP Collaboration at DESY in 1978
\cite{PLB76p361}, viz.\
the $\psi (3S)$, $\psi (2D)$, and  $\psi (4S)$,
lead to the $c\bar{c}$ interpretation of charmonium,
which predicts an in principle infinite number of excited $c\bar{c}$ states.
The observation of a variety of other charmonium states, with quantum numbers
different from $1^{--}$ \cite{PLB667p1}, supported this picture.

In recent years, several new charmonium(-like) resonances were reported
\cite{PRL89p102001,PRL91p262001,PRL93p072001,
PRL94p182002,PRL95p102003,PRL98p082001,PRL96p082003}.
Nevertheless, no new $c\bar{c}$ vector states have been discovered so far
\cite{ARXIV08082587}.
However, as we shall show here,
it has to be expected that,
with a relatively small improvement of the experimental statistics,
at least four, perhaps even five, new $\psi$ states
are within reach.
These should correspond to the substructures
in the cross sections of Ref.~\cite{PRL101p172001}, over the energy range
4.5--5.4 GeV, which we manage to extract from the data.

The $c\bar{c}$ vector states produced in $e^{+}e^{-}$ collisions,
can be observed in the production
cross sections of pairs of charmed hadrons.
Here, we shall combine the results of the RSE model for
meson-meson scattering and production
\cite{IJTPGTNO11p179,AP323p1215}
with the $\Lambda_{c}^{+}\Lambda_{c}^{-}$ data
of Ref.~\cite{PRL101p172001}
in order to identify possible signs of the
$\psi(5S)$, $\psi(4D)$, $\psi(6S)$, and $\psi(5D)$ states,
with masses predicted in Refs.~\cite{PRD21p772,PRD27p1527},
as well as an indication for the presence of the $\psi(3D)$ state
below the $\Lambda_{c}^{+}\Lambda_{c}^{-}$ threshold.

\begin{figure}[htbp]
\begin{center}
\begin{tabular}{c}
\includegraphics[height=180pt]{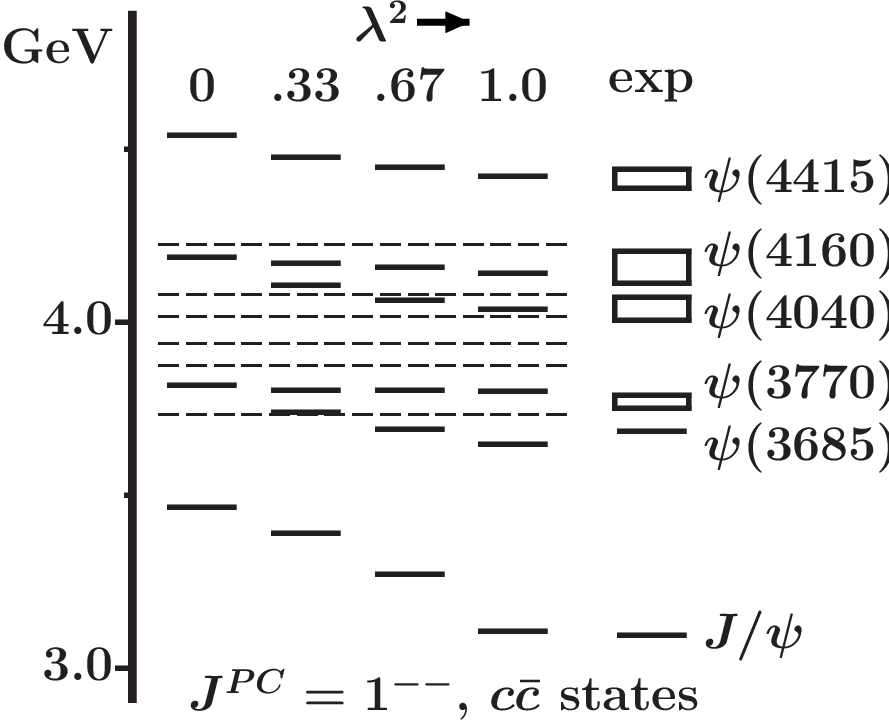}\\ [-10pt]
\end{tabular}
\end{center}
\caption[]{From the harmonic-oscillator (HO) spectrum to the charmonium
vector states \cite{PRD21p772}.
The parameter $\lambda$ represents the overall coupling of $c\bar{c}$
to the channels of open-charm meson pairs.
The relative couplings can be found in Ref.~\cite{ZPC21p291}.
On the left, under $\lambda^{2}=0$,
the HO states from Eq.~(\ref{Regge}) are shown,
for $m_{c}=1.562$ GeV and $\omega =0.19$ GeV.
Except for the ground state, the $L=0$ excitations
are degenerate with the $L=2$ ones.
Under $\lambda^{2}=0.33$, 0.67, and 1.0, it is shown how the model's
$\psi$ states develop towards the experimental spectrum \cite{PLB667p1},
which is given on the right.
The various dashed lines represent the thresholds for pairs of
$D$, $D^{\ast}$, $D_{s}$, and $D^{\ast}_{s}$ mesons.}
\label{Charming}
\end{figure}

In Refs.~\cite{PRD21p772,PRD27p1527} it was shown how the
bare $c\bar{c}$ spectrum
\begin{equation}
E_{nL}=2m_{c}+\omega\left( 2n+L+\frac{3}{2}\right)
\;\;\; ,
\label{Regge}
\end{equation}
turns into the physical spectrum of $c\bar{c}$ vector states
due to coupling to $D\bar{D}$,
$D\bar{D}^{\ast}$, $D^{\ast}\bar{D}^{\ast}$, $D_{s}\bar{D}_{s}$,
$D_{s}\bar{D}_{s}^{\ast}$ and $D_{s}^{\ast}\bar{D}_{s}^{\ast}$ meson loops.
This process is depicted, in a stepwise fashion, in Fig.~\ref{Charming}.

For $m_{c}=1.562$ GeV and $\omega =0.19 GeV$ \cite{PRD27p1527},
Eq.~(\ref{Regge}) gives
$E_{3}=E_{3,0}=E_{2,2}=4.549$ GeV,
$E_{4}=E_{4,0}=E_{3,2}=4.929$ GeV, and
$E_{5}=E_{5,0}=E_{4,2}=5.309$ GeV,
in the energy domain under consideration.
Within the RSE description for hadron-hadron scattering,
as depicted in Fig.~\ref{Charming},
we thus expect to find a $c\bar{c}$ $D$-wave resonance
about 20 MeV to 50 MeV below each of these values,
while the corresponding $S$-wave resonance comes out some 100 MeV
below the $D$-wave state.
In the case of $E_{3}=4.549$ GeV,
one has observed the $\psi (4S)(4415)$ resonance
at 4.4151$\pm$0.0079 GeV \cite{PLB667p1}.
However, the corresponding $\psi (3D)$ resonance,
which should have a mass of about 4.5 GeV,
has not yet been found.
For comparison,
Godfrey and Isgur predicted \cite{PRD32p189} the $3^{3}D_{1}$
$c\bar{c}$ vector state $\psi(3D)$ at a mass of 4.52 GeV.

In the present work,
we are going to explore a second feature of the bare states,
namely that at precisely the energy values $E_{nL}$ in Eq.~(\ref{Regge})
the RSE production cross sections exhibit sharp minima, i.e.,
approximate zeros, which are exact when there is no inelasticity.
In Ref.~\cite{IJTPGTNO11p179},
assuming quark-pair creation within a non-relativistic framework,
we found for
the $\ell$-th partial-wave propagator mode of strong interactions
the expression
(restricted to the one-channel case and
leaving out some parts not essential for our discussion here)
\begin{equation}
\bm{\Pi}_{\ell}(E)=\left\{
1-ij_{\ell}\left( pr_{0}\right)
h^{(1)}_{\ell}\left( pr_{0}\right)
\dissum{n=0}{\infty}
\fndrs{5pt}{\abs{g_{nL(\ell )}}^{2}}{-2pt}{E-E_{nL(\ell )}}
\right\}^{-1}
.
\label{propagator}
\end{equation}
The $q\bar{q}$ propagator of the RSE model
includes all the information
on virtual excitations of the $q\bar{q}$ system,
its real or virtual decay to hadron pairs,
and the hadron loops.
As in Ref.~\cite{IJTPGTNO11p179},
we take for the two-body scattering amplitude
the expression
\begin{equation}
\bm{T}_{\ell}(E)=\left\{ j_{\ell}\left( pr_{0}\right)^{2}
\dissum{n=0}{\infty}
\fndrs{5pt}{\abs{g_{nL(\ell )}}^{2}}{-2pt}{E-E_{nL(\ell )}}
\right\}
\bm{\Pi}_{\ell}(E)
\;\;\; .
\label{Tamplitude}
\end{equation}
Here, $E=\sqrt{s}$ represents
the total centre-of-mass (CM) energy,
$p$ the relative linear momentum in the CM frame,
$j_{\ell}$ and $h^{(1)}_{\ell}$
the order-$\ell$ spherical Bessel and Hankel functions of the first kind,
respectively , $L(\ell )$ the angular momentum
in the CM of the $q\bar{q}$ system
containing the flavours of the propagator,
$\ell$ the angular momentum in the hadron-hadron CM,
$r_{0}$ a distance parameter which we associate
with the average distance of quark-pair creation or annihilation,
and $g_{nL(\ell )}$ recoupling coefficients \cite{ZPC21p291}.
In the analysis below, we shall employ the fixed value
$r_{0}=3.8$ GeV$^{-1}$ ($\approx 0.76$ fm),
which is somewhat larger than what is usually used
in the RSE description of meson-meson interactions.
The RSE spectrum $E_{nL(\ell )}$ was given in Ref.~\cite{PRD27p1527}.

It is essential to notice that, for $E\to E_{nL(\ell )}$,
both the numerator and the denominator of expression (\ref{Tamplitude})
tend to infinity.
Hence, \bm{T} remains finite and, generally, non-zero.
However, the propagator (\ref{propagator})
goes to zero in this limit.

In Ref.~\cite{AP323p1215},
following a similar procedure as in Ref.~\cite{NPA744p127},
a relation between \bm{P} and \bm{T} was derived, reading
\begin{equation}
\bm{P}_{\ell}=
j_{\ell}\left( pr_{0}\right)
+i\,\bm{T}_{\ell}h^{(1)}_{\ell}\left( pr_{0}\right)
\;\;\; ,
\end{equation}
which, using Eqs.~(\ref{propagator}) and (\ref{Tamplitude}),
can also be written as
\begin{equation}
\bm{P}_{\ell}=
j_{\ell}\left( pr_{0}\right)
\bm{\Pi}_{\ell}(E)
\;\;\; .
\label{production}
\end{equation}
For the latter expression we find,
by the use of Eq.~(\ref{propagator}),
that the production amplitude of Eq.~(\ref{production})
goes to zero when $E\to E_{nL(\ell )}$.
This effect must be visible in experimental cross sections,
in particular for processes where two hadrons emerge,
as e.g.\ $e^{+}e^{-}\to\pi^{+}\pi^{-}$.
In the latter process, via the intermediate photon
and the creation of a pair of virtual current quarks,
the amplitude (\ref{production})
connects the electron-positron pair to the multi-hadron final states.

\begin{figure}[htbp]
\begin{center}
\begin{tabular}{c}
\scalebox{1.0}{\includegraphics{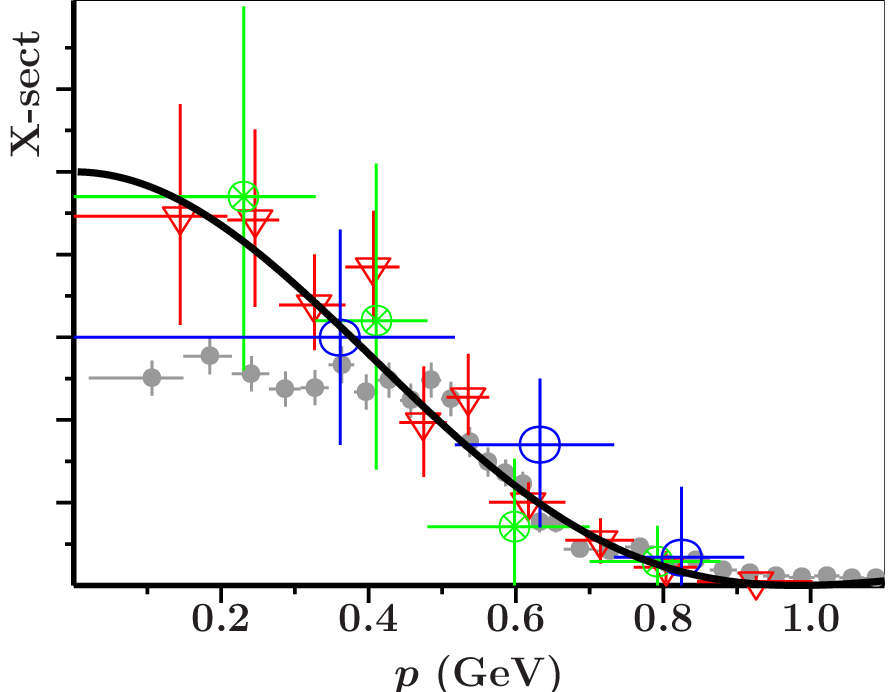}}\\ [-15pt]
\end{tabular}
\end{center}
\caption[]{\small
Experimental data
for $e^{+}e^{-}$ annihilation into baryon-antibaryon pairs, viz.\
$p\bar{p}$
(\definecolor{gray}{rgb}{0.6,0.6,0.6}{\color{gray}\bm{\bullet}})
\cite{PRD73p012005},
$\Lambda\bar{\Lambda}$
({\color{red}\bm{\bigtriangledown}}),
$\Sigma^{0}\bar{\Sigma}^{0}$
({\color{blue}\bm{\bigcirc}}),
and
$\Lambda\bar{\Sigma}^{0}$
({\color{green}\bm{\otimes}})
\cite{PRD76p092006}.
The $\Lambda\bar{\Lambda}$, $\Sigma^{0}\bar{\Sigma}^{0}$
and $\Lambda\bar{\Sigma}^{0}$ cross sections are scaled
with respect to the $p\bar{p}$ cross sections.
The curve (solid line)
is proportional to $\abs{j_{0}\left(pa\right)}^{2}$,
where $p$ represents the linear momentum and $a=3.2$ GeV$^{-1}$.
}
\label{BBbardata}
\end{figure}

The behaviour at threshold,
in particular for electron-positron annihilation
into baryon-antibaryon pairs dominantly in $S$-waves,
is in agreement with the data measured in experiment
\cite{PRD73p012005,PRD76p092006}.
According to the authors of
Refs.~\cite{IJMPA21p5565},
it is due to the cancellation of the phase-space factor
by the Coulomb form factor that the production cross sections for
$e^{+}e^{-}\to B\bar{B}$ do not vanish at threshold.
Alternatively, in the philosophy of Ref.~\cite{ARXIV08091149},
this may be due to the fact that initially only a pair of light current
quarks couples to the photon, for which phase space becomes practically
constant already right above threshold.
Even for a pair of current $c$ quarks phase space
varies less than 6\% in the here relevant invariant--mass--interval
from 4.6 GeV to 5.4 GeV.
In Fig.~\ref{BBbardata}, we have depicted,
as a function of linear momentum,
the production cross sections for $e^{+}e^{-}\to p\bar{p}$,
$\Lambda\bar{\Lambda}$, $\Lambda\bar{\Sigma}^{0}$, and
$\Sigma^{0}\bar{\Sigma}^{0}$.
We notice an excellent agreement between the proposed amplitude
of Eq.~(\ref{production}) and the data.
The deviation from the theoretical curve at lower momenta
for the proton-antiproton cross sections is probably
due to the presence of a nearby vector resonance below threshold.
Here, we shall study such a phenomenon for $\Lambda_{c}\bar{\Lambda}_{c}$.
\clearpage

When one of the produced particles
consists of heavy quarks and the others of light ones, then
some of the above-discussed zeros,
stemming from the heavy $q\bar{q}$ spectrum, should
be observable, as the production process most likely takes place via the
heavy $q\bar{q}$ propagator, and final-state interactions
between the heavy and light hadrons can be neglected.
For example, the non-resonant signal in
$e^{+}e^{-}\to\pi^{+}\pi^{-}\psi (2S)$
(see Fig.~5 of Ref.~\cite{PRL99p142002})
is divided into two substructures
\cite{PRD77p014033,PRD78p014032,PLB665p26},
since the full $c\bar{c}$ propagator (\ref{propagator}),
dressed with meson loops,
vanishes at $E_{3}=4.55$ GeV
\cite{PRD27p1527}.
In the same set of data, one may observe a lower-lying zero
at $E_{2}=4.17$ GeV
\cite{PRD27p1527},
more clearly visible in the data on
$e^{+}e^{-}\to\pi^{+}\pi^{-}J/\psi$
(see Fig.~3 of Ref.~\cite{PRL99p182004}).
The true $c\bar{c}$ resonances
can be found on the slopes
of the above-mentioned non-resonant structures
\cite{HEPPH0605317},
unfortunately with little statistical significance,
if any \cite{PRL95p142001}.

For the data of Ref.~\cite{PRL101p172001},
we thus find three zeros which are relevant.
In order to separate the resonance structure
of the $c\bar{c}$ vector states,
which we suppose to be mainly due to
the meson loops,
from the zeros of RSE,
we remove it from the $c\bar{c}$ propagator in such a way that
the zeros remain.
Thereto, inspired by Eq.~(2), we employ the
essentially phenomenological expression
\begin{equation}
A(p)=\fndrs{5pt}{j_{0}\left( pr_{0}\right)}{-15pt}
{1+
\fndrs{3pt}{e^\x{-2\left( pr_{0}\right)^{2}}}{-2pt}{r_{0}}\,
\left\{\dissum{n=0}{\infty}\,
\fndrs{2pt}{g_{n}}{-5pt}{\abs{E(p)-E_{n,0}}}
\right\}
}
\;\;\; ,
\label{amplitude}
\end{equation}
where  $p$ and $E(p)$ represent the linear momentum
in the CM system of the charmed-baryon pair
and their total invariant mass, respectively.
The only free parameter of expression (\ref{amplitude}), viz.\ $r_{0}$,
represents the average distance of light-quark-pair creation,
through which process we assume the charmed baryons
to be coupled to the $c\bar{c}$ system.

The expression in Eq.~(\ref{amplitude})
combines the following features:
\begin{enumerate}
\item
It displays zeros at $E_{n,0}=E_{n-1,2}$, like in the RSE expressions.
\item
By taking $\abs{E(p)-E_{n,0}}$,
instead of just $E(p)-E_{n,0}$ as in the RSE,
we avoid infinities (representing bound states) for real energies.
\item
It has no resonant structures.
\item
To lowest order, for which the denominator equals 1,
it represents the amplitude for a system which couples,
via a pair of current charm quarks and the RSE vertex,
to the photon.
The RSE vertex is given by a spherical Bessel function in the
CM frame of the outgoing pair of hadrons \cite{AP323p1215}.
\end{enumerate}
In practice, we shall here only consider the three zeros
$E_{3,0}$, $E_{4,0}$ and $E_{5,0}$, by setting $g_{n}=0$
for $n\geq 0$, except for $g_{3}$, $g_{4}$, and $g_{5}$.
The latter zeros
can be switched on ($g_{n=3,4,5}=1$) or off ($g_{n=3,4,5}=0$).
\clearpage

\begin{figure}[htbp]
\begin{center}
\begin{tabular}{c}
\includegraphics[height=320pt]{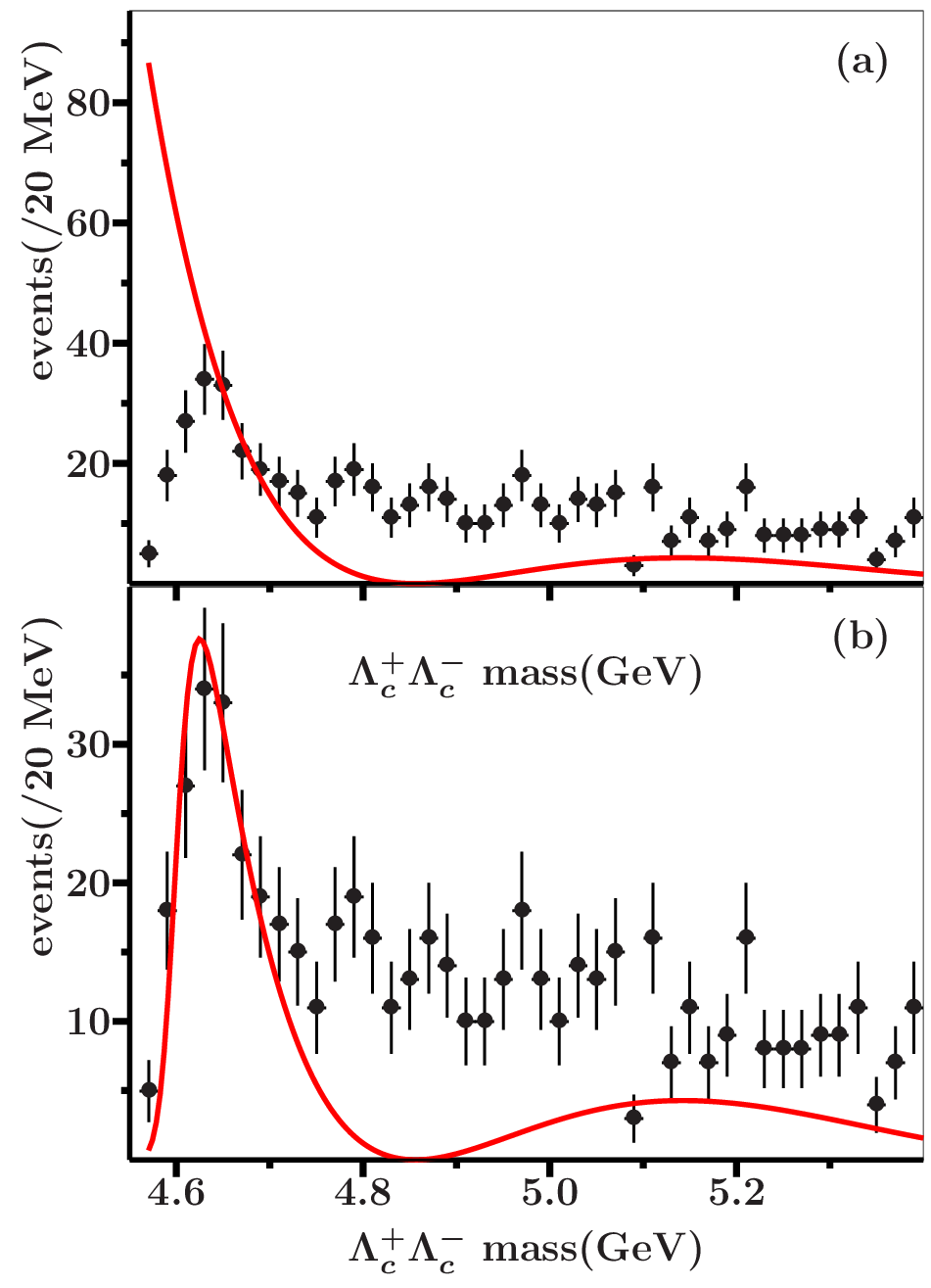}
\end{tabular}
\end{center}
\caption{Cross sections for
$e^{+}e^{-}\to\Lambda_{c}^{+}\Lambda_{c}^{-}$.
Comparison of the predictions from Eq.~(\ref{Xsect1})
for the case that the denominator of Eq.~(\ref{amplitude})
is not considered ($g_{3}=g_{4}=g_{5}=0$) (a),
with the prediction for the case that only the zero at 4.549 GeV
is taken into account ($g_{3}=1$, $g_{4}=g_{5}=0$) (b).
The data are taken from Ref.~\cite{PRL101p172001}.}
\label{wwz}
\end{figure}

In Fig.~\ref{wwz} we show the resulting cross sections, given by
\begin{equation}
\sigma = 2.0\,\alpha^{2}\pi r_{0}^{2}\abs{A}^{2}
\;\;\;\;\xrm{(events/20 MeV)}
\;\;\; ,
\label{Xsect1}
\end{equation}
for the case that we omit the denominator of Eq.~(\ref{amplitude}),
which represents the case of a structureless
$c\bar{c}$ propagator (Fig.~\ref{wwz}a),
and for the case that only the zero at 4.549 GeV
is taken into account (Fig.~\ref{wwz}b).
The result suggests that the structure near the
$\Lambda_{c}^{+}\Lambda_{c}^{-}$ threshold is caused
by the zero at $E_{3,0}=E_{2,2}=4.549$ GeV,
which implies the prediction of the $\psi(3D)$
about 20--50 MeV below that value.
\clearpage

We compare our results to the data of the Belle Collaboration
\cite{PRL101p172001}.
Upon further inspection of the data of Ref.~\cite{PRL101p172001}
(see also Fig.~\ref{wwz}),
we observe that the $\Lambda_{c}^{+}\Lambda_{c}^{-}$ channel,
via pion exchange \cite{PRD78p054022},
dominantly couples to $\Sigma_{c}\bar{\Sigma}_{c}$ channels.
We also use expression (\ref{Xsect1})
for the description of the corresponding cross section
in $\Lambda_{c}^{+}\Lambda_{c}^{-}$,
but now with $\abs{\bm{A}}^{2}$
replaced by \cite{AP323p1215}
\begin{equation}
\abs{
A\left( p_{\Lambda_{c}}\right)
+\begin{array}{c} \\ [-2pt]
\includegraphics[height=30pt]{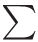}\\ [-2pt]
\Sigma_{c}\bar{\Sigma}_{c}\end{array}
f_{\Sigma_{c}\bar{\Sigma}_{c}\leftrightarrow\Lambda_{c}\bar{\Lambda}_{c}}\,
\fndrs{2pt}{A\left( p_{\Sigma_{c}}\right)}
{-5pt}{\sqrt{\mu_{\Sigma_{c}}/\mu_{\Lambda_{c}}}}
}^{2}
\;\;\; ,
\label{Xsect2}
\end{equation}
where $p_{\Sigma_{c}}$ and $\mu_{\Sigma_{c}}$ stand
for the linear momentum in the CM systems of
$\Sigma_{c}\bar{\Sigma}_{c}$
and the reduced mass of the two baryons, respectively.
The sum runs over the channels which,
assuming $J^{P}=\frac{3}{2}^{-}$ for $\Sigma_{c}(2800)$
\cite{PRL94p122002,PRD75p094017},
are given in Table~\ref{LSthresholds}.
\begin{table}[htbp]
\begin{center}
\begin{tabular}{||c||c|c||c||}
\hline\hline & & & \\ [-5pt]
{\small Channel} & {\small Threshold} & {\small Width} &
{\small $f_{\Sigma_{c}\bar{\Sigma}_{c}\leftrightarrow\Lambda_{c}\bar{\Lambda}_{c}}$}\\
& {\small (GeV)} & {\small (MeV)} & \\
& & & \\ [-6pt]
\hline\hline & & & \\ [-5pt]
$\Sigma_{c}(2455)\bar{\Sigma}_{c}(2455)$ &  4.907 & 5 & 0.040\\ [5pt]
$\Sigma_{c}(2520)\bar{\Sigma}_{c}(2520)$ &  5.036 & 30 & 0.019\\ [5pt]
$\Sigma_{c}(2455)\bar{\Sigma}_{c}(2800)$ &  5.252 & 77 & 0.032\\ [5pt]
\hline\hline
\end{tabular}
\end{center}
\caption[]{\small
$S$-wave thresholds
for selected pairs of charmed baryons \cite{PLB667p1}.
In the third column we indicate the sum of the widths
of the two charmed baryons, in order to have some idea of
the sharpness of the threshold.
The relative intensities of the contibutions
for the various channels are given in the fourth column.}
\label{LSthresholds}
\end{table}
\clearpage

\begin{figure}[htbp]
\begin{center}
\begin{tabular}{c}
\includegraphics[height=180pt]{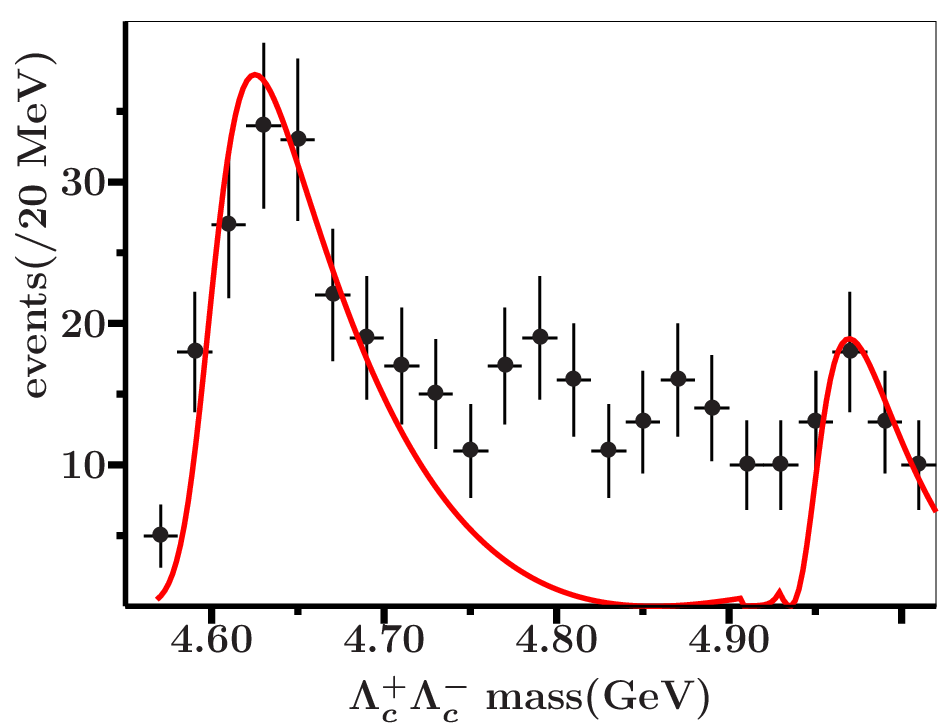}\\ [-175pt]
\hspace{180pt}{\bf (a)}\\ [160pt]
\includegraphics[height=180pt]{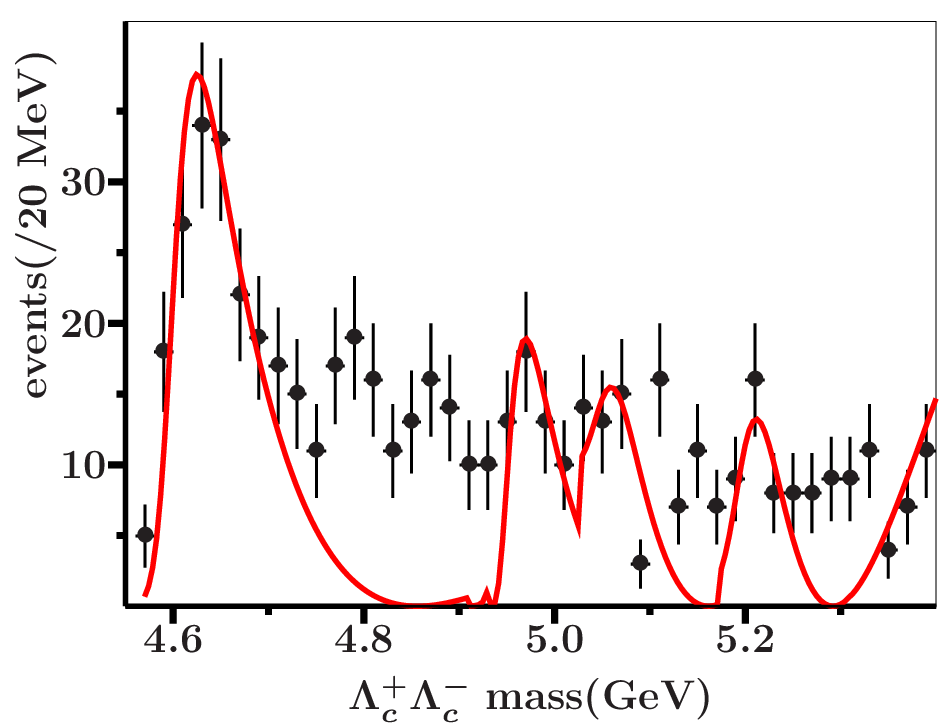}\\ [-175pt]
\hspace{180pt}{\bf (b)} \\ [140pt]
\end{tabular}
\end{center}
\caption{Opening of the
$\Sigma_{c}(2455)\bar{\Sigma}_{c}(2455)$,
$\Sigma_{c}(2520)\bar{\Sigma}_{c}(2520)$, and
$\Sigma_{c}(2455)\bar{\Sigma}_{c}(2800)$ channels
in the $\Lambda_{c}\bar{\Lambda}_{c}$ cross section:
detail (a), full interval (b).
The data are taken from Ref.~\cite{PRL101p172001}.}
\label{wwtz}
\end{figure}

The result, for $g_{3}=g_{4}=g_{5}=1$,
is shown in Fig~\ref{wwtz}.
Notice that the $\Sigma_{c}(2455)\bar{\Sigma}_{c}(2455)$ channel
does not start out at threshold, namely 4.907 GeV,
but is suppressed up to the zero at 4.929 GeV,
which seems to agree with experiment.
In the absence of the latter zero,
all signals due to the opening of new channels
would be sharply peaked at threshold.
\clearpage

\begin{figure}[htbp]
\begin{center}
\begin{tabular}{c}
\includegraphics[height=200pt]{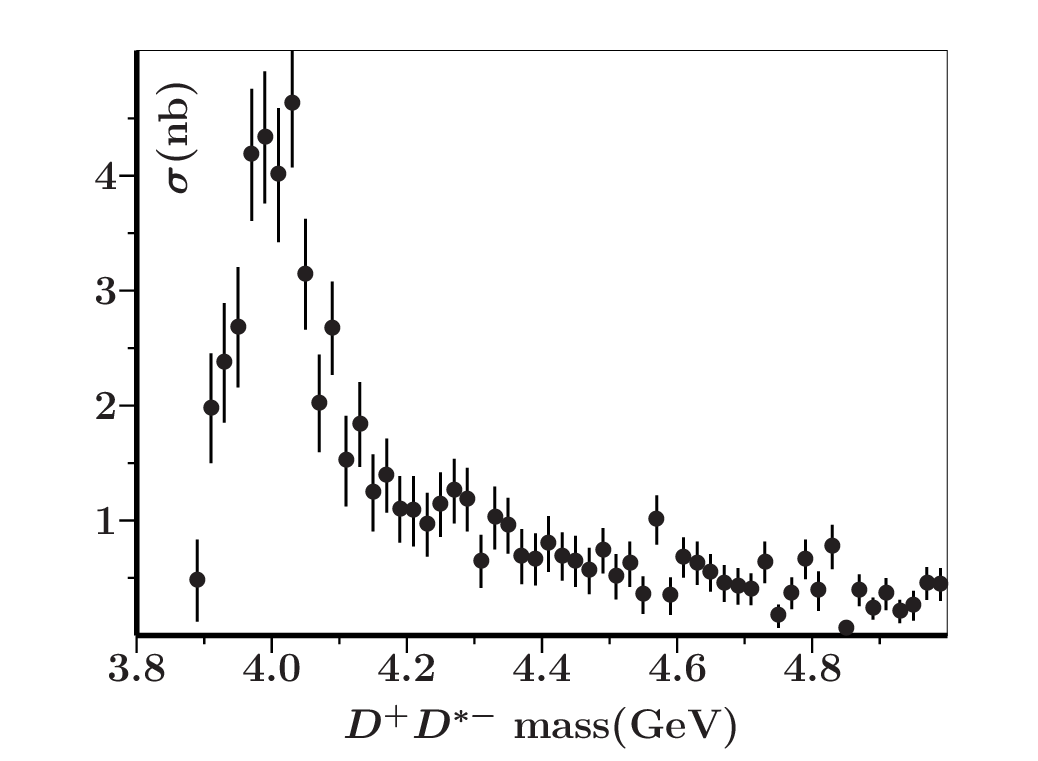}\\ [-20pt]
\end{tabular}
\end{center}
\caption{Previous $e^{+}e^{-}\to D^{+}D^{\ast -}$ data from
the Belle Collaboration \cite{PRL98p092001}.}
\label{HEPEX0608018}
\end{figure}

Since no further $\Sigma_{c}\bar{\Sigma}_{c}$ thresholds are known
to open in the invariant-mass interval
from 4.57 Gev ($\Lambda_{c}\bar{\Lambda}_{c}$ threshold)
to 4.91 GeV ($\Sigma_{c}(2455)\bar{\Sigma}_{c}(2455)$ threshold)
(see Fig.~\ref{wwtz}a),
we must conclude that the remaining structures
represent two possible resonances, viz.\
at about $4.79$ GeV and $4.87$ GeV,
some 140 MeV and 60 MeV below the zero at $E_{4}=4.929$ GeV, respectively.
This is right at the positions where, with the RSE model for $c\bar{c}$
states, one expects to find the $\psi (5S)$ and $\psi (4D)$, respectively.
Similar structures, but also with low statistics,
can be observed in the data for $e^{+}e^{-}\to D^{+}D^{\ast -}$
published by the Belle Collaboration in
Fig.~2b of Ref.~\cite{PRL98p092001}
(see Fig.~\ref{HEPEX0608018}).

Some remarks are due with respect to the theoretical curve
of Fig.~\ref{wwtz}b.
Since the $\Sigma_{c}(2520)$ has a width of about 15 MeV,
the $\Sigma_{c}(2520)\bar{\Sigma}_{c}(2520)$ channel
will effectively open below 5.036 GeV.
We have accounted for that by choosing threshold 10 MeV lower,
which seems to better agree with the data.
Then, the width of the $\Sigma_{c}(2800)$ is roughly 75 MeV, with
a large error.
We found that the data are best described by choosing
the $\Sigma_{c}(2455)\bar{\Sigma}_{c}(2800)$ channel to open
at 5.172 GeV.
\clearpage

\begin{figure}[htbp]
\begin{center}
\begin{tabular}{c}
\includegraphics[width=400pt]{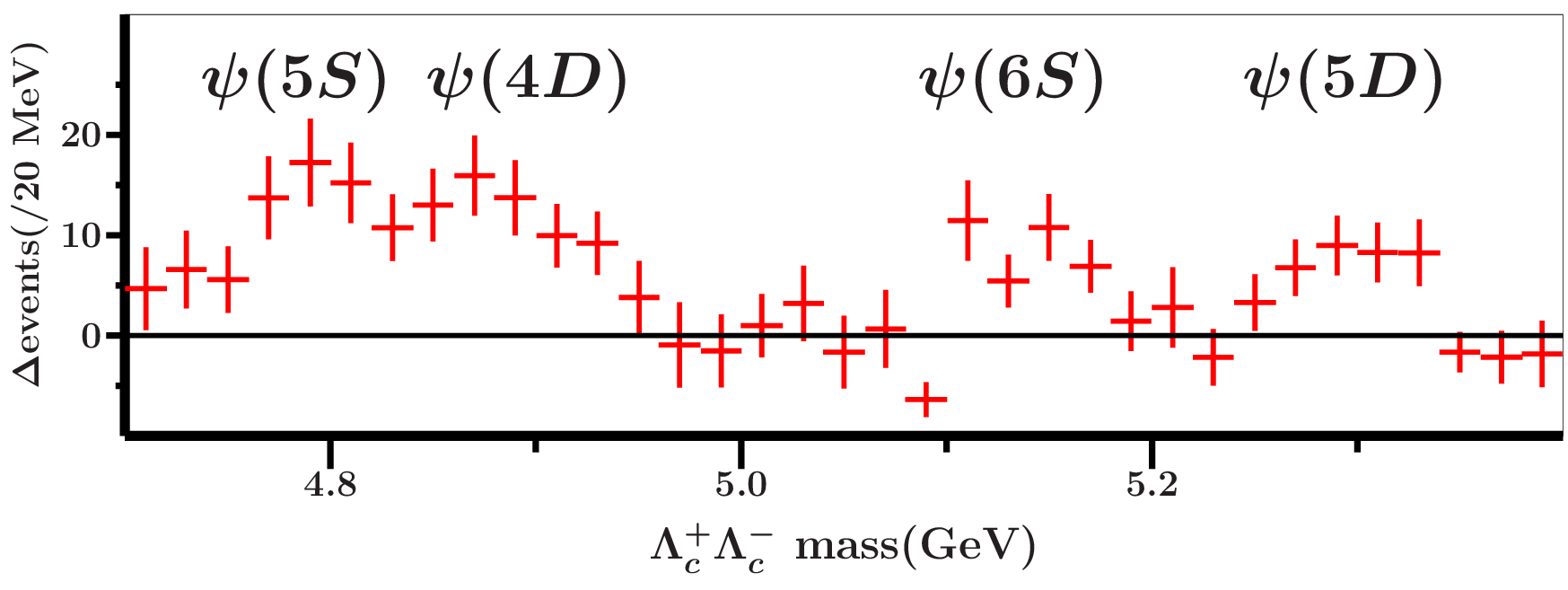}\\ [-20pt]
\end{tabular}
\end{center}
\caption{Difference between the data of Ref.~\cite{PRL101p172001}
and the theoretical curve of Fig.~\ref{wwtz}.}
\label{Difference}
\end{figure}

We interpret the theoretical curve of Fig.~\ref{wwtz}
as the non-resonant signal in $\Lambda_{c}\bar{\Lambda}_{c}$.
The remaining structures, which are shown in Fig.~\ref{Difference},
may stem from $c\bar{c}$ vector resonances.

We clearly observe the $\psi (5S)$ and $\psi (4D)$
in Fig.~\ref{Difference},
with 3--4$\sigma$,
at $\approx 4.79$ GeV and $\approx 4.87$ GeV.
Less clearly, with 1--2$\sigma$,
we see two more indications for resonant structures, namely
at $\approx 5.13$ GeV and $\approx 5.29$ GeV,
about 180 MeV and 20 MeV
below the RSE zero at $E_{5}=5.309$ GeV, respectively.
These might be associated with
the $\psi (6S)$ and $\psi (5D)$ charmonium states, respectively.
However, more statistics is definitely needed
for confirmation.

Summarising, the near-threshold enhancement in
$e^{+}e^{-}\to\Lambda_{c}\bar{\Lambda}_{c}$
observed by the BELLE collaboration \cite{PRL101p172001} is explained here as
the combined effect of a normal threshold behaviour and a sub-threshold zero
in the amplitude.  Furthermore, we conclude that the data
\cite{PRL101p172001,PRL99p142002,PRL99p182004}
confirm the zeros of the $c\bar{c}$ propagator
which 25 years ago were predicted in Ref.~\cite{PRD27p1527}.

Based on their properties,
we have found some indication for 4
not very broad (30--60 MeV) new $\psi$ states
in the BELLE data \cite{PRL101p172001}:
the $\psi (5S)$ at $\approx 4.79$ GeV,
the $\psi (4D)$ at $\approx 4.87$ GeV,
the $\psi (6S)$ at $\approx 5.13$ GeV,
and the $\psi (5D)$ at $\approx 5.29$ GeV.
Moreover,  we also see an indirect indication for the existence
of the $\psi (3D)$ at $\approx 4.50$ GeV,
from the threshold behaviour of $\Lambda_{c}\bar{\Lambda}_{c}$.

We suggest that the two structures existing in
the invariant-mass interval 4.7--4.9 GeV,
possibly the $\psi(5S)$ and $\psi(4D)$ states,
as well as the other two structures in
the invariant-mass interval 5.1--5.3 GeV,
possibly the $\psi(6S)$ and $\psi(5D)$ states,
are searched for in future experiments.
A possible confirmation of further zeros in the $c\bar{c}$ propagator,
given by Eq.~(\ref{Regge}), would also be of great relevance.
\vspace{10pt}

We are grateful to Kanchan Khemchandani for
many useful suggestions
This work was supported in part by
the \emph{Funda\c{c}\~{a}o para a Ci\^{e}ncia e a Tecnologia}
\/of the \emph{Minist\'{e}rio da Ci\^{e}ncia, Tecnologia e Ensino Superior}
\/of Portugal, under
contract POCI/\-FP/\-81913/\-2007 and fellowship SFRH/BPD/34819/2007.
One of us (X.L.) would like to thank the National Natural Science
Foundation of China for grant 10705001.

\newcommand{\pubprt}[4]{{#1 {\bf #2}, #3 (#4)}}
\newcommand{\ertbid}[4]{[Erratum-ibid.~{#1 {\bf #2}, #3 (#4)}]}
\def\AP{Ann.\ Phys.}
\def\IJMPA{Int.\ J.\ Mod.\ Phys.\ A}
\def\IJTPGTNO{Int.\ J.\ Theor.\ Phys.\ Group Theor.\ Nonlin.\ Opt.}
\def\NPA{Nucl.\ Phys.\ A}
\def\PLB{Phys.\ Lett.\ B}
\def\PRD{Phys.\ Rev.\ D}
\def\PRL{Phys.\ Rev.\ Lett.}
\def\ZPC{Z.\ Phys.\ C}

\end{document}